\documentclass{svjour3}
\usepackage{fix-cm}
\usepackage{amsmath}
\def\be{\begin{equation}}
\def\ee{\end{equation}}
\def\bq{\begin{eqnarray}}
\def\eq{\end{eqnarray}}

\journalname{General Relativity and Gravitation}

\begin{document}

\title{Generalized Euclidean  stars with equation of state}

\author{ G. Z. Abebe \and S. D. Maharaj  \and  K. S. Govinder}

\institute{
G. Z. Abebe \and S. D. Maharaj  \and  K. S. Govinder \at
Astrophysics and Cosmology Research Unit,
 School of Mathematics, Statistics and Computer  Science,
 University of KwaZulu-Natal, Private Bag X54001, Durban 4000, South Africa\\ \\
\\ \\
\email{maharaj@ukzn.ac.za} \\ \\
}
\date{Received: date / Accepted: date}

\maketitle

\begin{abstract}
 We consider the general case of an accelerating, expanding and shearing model of a radiating relativistic star using Lie symmetries.  We obtain the  Lie symmetry generators that leave the  equation for the junction condition  invariant,  and find  the Lie algebra  corresponding to the optimal system of the symmetries. The symmetries  in the optimal system  allow us to transform the boundary condition  to  ordinary differential equations.  The various cases for which the resulting systems of equations can be solved are identified. For each of these cases the boundary condition is integrated and the gravitational potentials are found explicitly. A particular group invariant solution produces  a class of models which contains Euclidean stars as a special case. Our generalized model satisfies  a linear equation of state in general. We thus establish a group theoretic basis for our generalized model with an equation of state. By considering  a particular example we show that the weak, dominant and strong  energy conditions are satisfied. 

\end{abstract}

\section{Introduction \label{into} }
Models of relativistic radiating stars with expansion, shear and acceleration are  the most general  in spherically symmetric spacetimes. They are  important in investigating  physical features including stellar stability, surface luminosity, relaxational  effects, causal temperature  profiles, particle production at the surface,   the cosmic censorship hypothesis of Penrose,  and  gravitational collapse. The exterior Vaidya \cite{1} spacetime must match with the interior spacetime of the  radiating star leading to the generation  of  the junction surface condition relating the radial pressure to the heat flux. This junction condition was  obtained  by Santos \cite{2} for the first time.   The first exact solution for a relativistic  radiating star with shear was obtained by Naidu et al. \cite{15}. Their treatment has been generalized by   Rajah and Maharaj \cite{16} by treating the junction condition as a Riccati equation. Thirukkanesh and Maharaj \cite{10a}  generated two classes of exact solutions that contain all previously known models as  special cases. Recently Abebe et al. \cite{10ab} generated exact solutions using the Lie symmetry approach and obtained metrics in terms of elementary functions. This class of models has the feature of containing certain solutions of the traveling  wave type and   the others that are expressed in terms of  a self-similar variable. All the above mentioned treatments were modeled for radiating stars where the fluid particles  are traveling with geodesic motion and regained the Friedmann dust model in the absence of heat flux in the relevant limit. The presence of acceleration  changes the nature of the model.
  
 The presence of acceleration with shear increases the complexity of the model substantially.  A systematic study  of these models was initiated by Chan \cite{chan1}  for a configuration that is initially static  and then collapses. In several analyses Chan \cite{G4,G33,H4} and   Pinheiro  and Chan \cite{GR,mod}  studied  the   luminosity, viscous effects and other physical features  in the presence of shear.  The physical analysis  in these  treatments \cite{chan1,G4,G33,H4,GR,mod} were accomplished using a numerical approach. Recently Thirukkanesh et al. \cite{G3} obtained exact solutions for an accelerating and expanding  model with shear by transforming the junction condition into linear, Bernoulli and  inhomogeneous Riccati equations. 
 Herrera and Santos \cite{G4a} studied the physical features in general when anisotropy  is present. They also introduced the idea of Euclidean stars \cite{G5} which are also modeled  with nonvanishing shear. Such stars  were further analysed by Govender et al. \cite{G6}  and Govinder and Govender \cite{G7}. The effect of shear in a radiating star undergoing dissipative collapse has been recently studied by Govender et al. \cite{mks}. The presence of acceleration and the effect of shear change the nature of    the resulting junction condition equation.

  The aim of this paper is to generate  exact solutions to the boundary condition equation of a radiating star by applying group theory to the relevant differential equations. In previous investigations  the Lie theory of extended groups applied to differential equations  has been used very successfully to  generate exact solutions to the Einstein field equations. Such treatments include Govinder et al. \cite{a17}, Leach and Govinder \cite{b17}, Hansraj et al. \cite{c17}, Msomi et al. \cite{d17,f17,g17}, Kweyama et al. \cite{e17}, Govinder  and Hansraj \cite{h17}. The boundary condition in Euclidean stars has been solved by Govinder and Govender \cite{G7} with the help of a particular Lie symmetry. Their treatment is believed to be the first Lie symmetry approach for a radiating general relativistic star.   Exact solutions have also been found  for a shearing model when the fluid particles are  traveling with geodesic motion (Abebe et al. \cite{10ab}) and for conformally flat metrics (Abebe et al. \cite{i17}) using the Lie approach.  Motivated by these treatments,  we analyze    the boundary condition of the general case with acceleration, expansion and   shear using a Lie symmetry approach. This general treatment is likely to produce features not present in earlier studies.
 
We present the junction condition for an accelerating, expanding and shearing radiating star in   Sect. \ref{5sec1}. This is a highly nonlinear partial differential equation in the metric functions. We obtain the Lie point symmetries  for the junction  condition and  generate  the optimal system of these symmetries.  Using  a particular  symmetry in the optimal system, we transform  the boundary condition to  an ordinary differential equation and obtain group invariant solutions in Sect. \ref{xzx}.  We find  new classes of exact solutions to the boundary condition and regain known solutions. We show that  the new solutions found satisfy a barotropic equation   of state in Sect. \ref{physi}. In Sect. \ref{energy} we consider the physical features and analyze the energy conditions in a particular example.  In Sect. \ref{5sec4} we make concluding remarks. The remaining cases of the optimal system for which the boundary condition can be integrated are listed in the Appendix in Sect. \ref{Appen}

  \section{The model\label{5sec1}}
  We consider the most general form of a spherically symmetric radiating star with acceleration, expansion and shear.   The  line element for the interior spacetime of such a star is given by
 \begin{equation}\label{5a1}
 ds^2=-A^2dt^2+B^2  dr^2+Y^2 \left( d\theta^2+\sin ^2\theta d\phi^2 \right), 
 \end{equation}
 where the metric functions $A$, $ B$ and $Y$  are functions of the coordinate radius $r$ and the temporal time variable $t$.   The fluid four-velocity   \textbf{u} is comoving and is given by $u^a=\frac{1}{A}\delta^a_0
 $. The kinematical quantities, the acceleration $\dot{u}^a$, the expansion scalar $\Theta$,  and the magnitude of the shear scalar $\sigma$,  are give by
\begin{subequations}
\begin{eqnarray}
\dot{u}^a&=&\left(0,\frac{A_r}{AB^2},0,0 \right), \\
\Theta &=& \frac{1}{A}\left(\frac{B_t}{B}+2\frac{Y_t}{Y} \right), \\
\sigma &=&-\frac{1}{3A}\left( \frac{B_t}{B}-\frac{Y_t}{Y}\right), 
\end{eqnarray} 
\end{subequations}
where the subscripts denote  differentiation with respect to $r$ and $t$.
 The energy momentum tensor for the shearing  model has the form
 \begin{equation}\label{7aa}
 T_{ab}=\left(\mu+p \right) u_au_b+pg_{ab}+q_au_b+q_bu_a+\pi_{ab},
 \end{equation}
where $\mu$ is the density, $p$ is the isotropic pressure, $q_a$ is the heat flux, and $\pi_{ab}$ is the anisotropic stress.  The quantity \textbf{u} is the fluid four-velocity which satisfies $u^au_a=-1$ and $u^a=\frac{1}{A}\delta ^a_0$. The stress tensor is given by
\be 
\pi_{ab}=\left(  p_{\parallel}-p_{\perp}\right)\left(n_an_b-\frac{1}{3}h_{ab} \right). 
\ee
We have introduced the radial pressure $p_{\parallel}$, the tangential pressure $p_{\perp}$,  the projection tensor $h_{ab}$  and a unit radial vector  \textbf{n} given by $n^a=\frac{1}{B}\delta ^a_1$.
The radial and the tangential pressures give  the isotropic pressure 
$ 
p=\frac{1}{3}\left( p_{\parallel}+2p_{\perp} \right)
$. Since the heat must  flow in the radial direction  the heat flow  vector \textbf{q} may be written as  
\be
  q^a=(0,Bq,0,0),
  \ee
and $q^au_a=0$.  
  
The Einstein field equations for the interior of the star have the form
\begin{subequations}\label{145}
 \begin{eqnarray}
 \mu&=&\frac{2}{A^2}\frac{B_t}{B}\frac{Y_t}{Y}+\frac{1}{Y^2} +\frac{1}{A^2}\frac{Y_t^2}{Y^2}-\frac{1}{B^2}\left(2 \frac{Y_{rr}}{Y}+\frac{Y_r^2}{Y^2}-2\frac{B_r}{B}\frac{Y_r}{Y}\right), \label{mu}  \\
 p_ \parallel&=&\frac{1}{A^2}\left( -2\frac{Y_{tt}}{Y}-\frac{Y_t^2}{Y^2}+2\frac{A_t}{A}\frac{Y_t}{Y}\right)+\frac{1}{B^2}\left( \frac{Y_r^2}{Y^2}+2\frac{A_r}{A}\frac{Y_r}{Y}\right)-\frac{1}{Y^2},\label{14p} \\
 p_ \perp&=&-\frac{1}{A^2}\left( \frac{B_{tt}}{B}-\frac{A_t}{A}\frac{B_{t}}{B}+\frac{B_t}{B}\frac{Y_t}{Y}-\frac{A_t}{A}\frac{Y_t}{Y}+\frac{Y_{tt}}{Y}\right)\nonumber\\
&&+\frac{1}{B^2}\left( \frac{A_{rr}}{A}-\frac{A_r}{A}\frac{B_r}{B}+\frac{A_r}{A}\frac{Y_{r}}{Y}-\frac{B_r}{B}\frac{Y_r}{Y}+\frac{Y_{rr}}{Y}\right),  \label{14r}\\
 q&=&-\frac{2}{AB}\left(-\frac{Y_{rt}}{Y} +\frac{B_t}{B}\frac{Y_r}{Y}+\frac{A_r}{A}\frac{Y_t}{Y}\right),\label{14d} 
 \end{eqnarray}
 \end{subequations}
for the metric \eqref{5a1}. The matter variables  $\mu$, $p_\parallel$, $p_\perp$ and $q$ can be determined explicitly once the potential functions $A$, $B$ and $Y$ are known. 
Equations \eqref{145} describe the  gravitational interactions in the  interior of an accelerating, expanding and shearing star with heat flux and anisotropic pressure.

The surface of a  spherically symmetric radiating star is the boundary between  the interior  and the exterior spacetimes. The interior spacetime \eqref{5a1} has to be matched at the surface of the star to the exterior Vaidya spacetime
\be\label{va1}
ds^2=-\left(1-\frac{2m(v)}{R} \right)dv^2-2dvdR+R^2\left( d\theta^2+\sin ^2\theta d\phi^2 \right). 
\ee
In \eqref{va1} the function  $m(v)$  is the mass of the star  at infinity. The metrics \eqref{5a1} and \eqref{va1}  have to be matched  at the boundary of the star. The matching of the metrics   and  the  extrinsic curvature at the surface of the star  give  the  junction conditions
\begin{subequations}\label{rev1}
\begin{eqnarray}
Adt&=& \left(1-\frac{2m}{R_\Sigma}+2\frac{dR_\Sigma}{dv} \right)^\frac{1}{2}dv, \\
\left(Y \right)_{\Sigma}&=&R_{\Sigma }(v),\\
m(v)&=&\left[\frac{Y}{2}\left(1 +\frac{Y_t^2}{A^2}-\frac{Y_r^2}{B}\right)  \right] _{\Sigma}, \\
(p_ \parallel)_{\Sigma}&=&(q)_{\Sigma},\label{santo}
\end{eqnarray}
\end{subequations}
at the hypersurface  $\Sigma $   of the radiating sphere. The junction condition \eqref{santo}, for shear-free spacetimes was first found by Santos \cite{2}, and, later extended to shearing  spacetimes  by Glass \cite{11a}. Equation \eqref{santo}, together with \eqref{14p}  and \eqref{14d}, leads to the junction condition equation
\begin{eqnarray} \label{c51} 
&&2AB^2YY_{tt}+AB^2Y_t^2-2B^2YA_tY_t-2ABYA_rY_t+2A^2BYY_{rt}-2A^2YA_rY_r\nonumber\\
&&-2A^2YB_{t}Y_{r}-A^3Y_r^2+A^3B^2=0,
\end{eqnarray}
 at the boundary of the star. Equation \eqref{c51} determines the gravitational behaviour of the radiating anisotropic star with nonzero shear, acceleration and expansion. We need to solve \eqref{c51} exactly to complete the model. This equation is a highly nonlinear partial differential equation and difficult to solve directly. Therefore we undertake a group theoretic analysis of \eqref{c51} in order to find useful solutions.

A differential equation of order $k$
\begin{eqnarray}
& F& (r,t,A,B,Y,A_r,B_r,Y_r,A_t,B_{t}, Y_t, \nonumber \\
&& A_{rr},B_{rr},Y_{rr},A_{rt},B_{rt},Y_{rt},A_{tt},B_{tt},Y_{tt}, \dots ) =0,
\end{eqnarray}
where $A=A(t,r)$, $B=B(r,t)$ and $Y=Y(r,t)$ admits a Lie point symmetry of  the form
\bq \label{sym10}
X&=&\xi_1\frac{\partial}{\partial r}+\xi_2\frac{\partial}{\partial t}+\eta_1\frac{\partial}{\partial A}+\eta_2\frac{\partial}{\partial B}+\eta_3\frac{\partial}{\partial Y},
\eq   
 where $\xi_i$ and $\eta_j$ are  functions of $r,t,A,B$ and $Y$  ($i=1,2$  and  $j=1,2,3$), provided that
\be
\left. X^{[k]}F\right|_{F=0}=0, 
\ee
where $X^{[k]}$ is the $k$th prolongation of the symmetry $X$. The process is algorithmic. Using  the computer software package PROGRAM LIE \cite{114} we find that   \eqref{c51} possesses the symmetries
\begin{subequations}\label{c5sym}
\begin{eqnarray}
X_1&=&A\frac{d\beta }{dt}\frac{\partial}{\partial A}-\beta (t)\frac{\partial}{\partial t},\label{c5sym2}\\
X_2&=&B\frac{d\alpha}{dr}\frac{\partial}{\partial B}-\alpha(r) \frac{\partial}{\partial r},\label{c5sym22}\\
X_3&=&A\frac{\partial}{\partial A}+B\frac{\partial}{\partial B}+Y\frac{\partial}{\partial Y},\label{c5sym3}
\end{eqnarray}
\end{subequations}
where $\beta (t)$ and $\alpha(r) $ are nonzero arbitrary functions of $t$ and $r$ respectively. This reveals that equation \eqref{c51} is invariant under scalings of $A$ and $t$ together, $B$ and $r$ together,  and $A$, $B$ and $Y$ together. Note that, since $\beta (t)$ and $\alpha(r) $ are arbitrary functions,  they mask the expected  invariance under translations in $t$ and $r$ separately (which is obtained by setting those functions to constants). These symmetries  will be used to generate group invariant solutions. 

Group invariant solutions obtained  by using any linear combination  of the individual symmetries in  \eqref{c5sym} may be transformed to  the symmetries in the optimal system \cite{113}. 
 We determine an optimal system of \eqref{c5sym} to be 
\begin{subequations}\label{mar7}
\begin{eqnarray}
X_1&=&A\frac{d\beta }{dt}\frac{\partial}{\partial A}-\beta (t)\frac{\partial}{\partial t},\\
aX_1+X_3&=&\left( a\frac{d\beta }{dt}+1\right) A\frac{\partial}{\partial A}+B\frac{\partial}{\partial B}+Y\frac{\partial}{\partial Y}-a\beta (t)\frac{\partial}{\partial t},\\
X_2-bX_1&=&-bA\frac{d\beta }{dt}\frac{\partial}{\partial A}+B\frac{d\alpha}{dr}\frac{\partial}{\partial B}-\alpha(r) \frac{\partial}{\partial r}+b\beta (t)\frac{\partial}{\partial t}.
\end{eqnarray}
\end{subequations}

\section{Generalized Euclidean stars}\label{xzx}
 The symmetries in \eqref{mar7} may  be applied to reduce the governing partial differential equation to ordinary differential equations using the infinitesimal  generators obtained via the Lie approach  \cite{113,112}.  A number of different cases are possible which lead to exact solutions of  the boundary condition \eqref{c51}. In this section  we present only the most interesting case which has physically applicable features. The remaining integrable cases are included in the Appendix in Sect. \ref{Appen}.
 
The invariants of the particular generator
\be \label{traveling}
X_2-bX_1=-bA\frac{d\beta }{dt}\frac{\partial}{\partial A}+B\frac{d\alpha}{dr}\frac{\partial}{\partial B}-\alpha(r) \frac{\partial}{\partial r}+b\beta (t)\frac{\partial}{\partial t},
\ee
 can be found from the system
 \begin{equation}
\frac{dt}{b\beta (t)}=\frac{dr}{-\alpha (r)}=\frac{dA}{-bA\frac{d\beta }{dt}}=\frac{dB}{B\frac{d\alpha}{dr}}=\frac{dY}{0}
\end{equation}
to be
\begin{subequations}\label{inv52}
\begin{eqnarray}
x&=&\int\frac{dt}{b\beta (t)}+\int\frac{dr}{\alpha (r) } \label{trav11},\\
A&=&\frac{f(x)}{\beta (t)},\\
B&=&\frac{g(x)}{\alpha (r)},\\
Y&=&h(x).
\end{eqnarray}
\end{subequations}

 We note that solutions obtained  via the symmetry \eqref{traveling}  contain traveling waves. This  can be seen explicitly  if  we set  $\beta (t) = \alpha (r)=1$. Then the independent variable \eqref{trav11} becomes 
\be 
x=\frac{1}{b}t+r,
\ee
  with  wave  speed $1/b$. This fact arises  since the generator \eqref{traveling} gives the reduced form  \be \label{wave}
 X_2-bX_1=b\frac{\partial}{\partial t}-\frac{\partial}{\partial r},
 \ee
 in this case.

For  the transformation \eqref{inv52}, equation \eqref{c51}   reduces to 
\be \label{ss2}
 g'+\left(\frac{f'}{f}-\frac{h''}{h'}\right)g-\left(\frac{b^2 f^3-2 h f' h'+f \left(h'^2+2 h h''\right)}{2 b f^2 h h'}\right)g^2 +\left(b f'+\frac{b f h'}{2 h}\right)=0,
 \ee
 which is a nonlinear equation in the functions $f$, $g$ and $h$. Primes denote differentiation with respect to the variable $x$. We  have expressed \eqref{ss2}  in a form which can be interpreted as a Riccati equation in $g$. In the Appendix we undertake a more detailed analysis of this equation.  Here we note that the equation can be simplified  if  we assume 
\be \label{sew01}
f=cg,
\ee
 where $c$ is an arbitrary constant. Then \eqref{ss2} becomes
\be\label{berno}
g'+\left(\frac{(c b-1) h'^2-2 h h''}{2 (c b+1) h h'}\right)g-\left(\frac{c^2 b^2}{2 (c b+1)^2 h h'}\right) g^3=0,
\ee
which is a Bernoulli equation  in $g$ of degree three. Even though $h$ is an unknown function we can integrate \eqref{berno} to obtain 
\be \label{sew3}
g(x)=\frac{ h^{\frac{1-c b}{2 (cb+1)}} h'^{\frac{1}{cb+1}}}{\left( d - \frac{c^2 b^2 }{ (cb+1)^2 } \int_1^x  h(z)^{\frac{-2c b}{cb+1}}h'(z)^{\frac{1-cb}{cb+1}} \, dz\right)^{1/2} },
\ee
where $d$ is a constant of integration. Thus we have found an exact solution to the boundary condition \eqref{c51}. The solution is expressible in terms of the function $h$ which is arbitrary.
Thus the potential functions become
\begin{subequations}\label{fff1}
\begin{eqnarray}
A&=&\frac{c h(x)^{\frac{1-c b}{2 (cb+1)}} h'(x)^{\frac{1}{cb+1}}}{\beta (t)\left( d - \frac{c^2 b^2 }{ (cb+1)^2 } \int_1^x  h(z)^{\frac{-2c b}{cb+1}}h'(z)^{\frac{1-cb}{cb+1}} \, dz\right)^{1/2} },\\
B&=&\frac{\beta (t)}{c \alpha (r)}A, \quad Y \space = \space h(x),
\end{eqnarray}
\end{subequations}
 where $h$ is  a function of $x= \int\frac{dt}{b\beta (t)}+\int\frac{dr}{\alpha (r) }$. This solution is expressed in terms of arbitrary constants and the arbitrary functions  $\alpha (r)$, $\beta(t)$ and $h(x)$ so that we can generate infinitely many solutions to the boundary condition for particular choices.
 
 For physical applications we need to show that the integrals in \eqref{fff1} may be  written in closed form. This is possible for particular choices of the function $h(x)$. As a first example we set
 \be
 h(x)=\exp\left(e+kx \right),
 \ee
 where $e$ and $k$ are constants. Then the potentials \eqref{fff1} become
 \begin{subequations}\label{vanishing}
\begin{eqnarray}
A&=& ck^{\frac{1}{1+c b}}\exp \left( \frac{\left( 3-c b\right) (e+k x)}{2+2c b}\right)  \frac{1}{\beta (t)}    \left(d-\frac{c^2 b^2}{(1+c b) (3 c b-1)k^{\frac{2 c b}{1+c b}} }  \right. \nonumber \\
&& \left. \left[\exp\left( \frac{ (e+k) (1-3c b) }{1+c b}\right) -\exp \left( \frac{(e+k x) (1-3c b)}{1+c b}\right) \right] \right)^{-1/2},\\
B&=&\frac{\beta (t)}{c \alpha (r)}A, \quad Y=\exp\left(e+kx \right)  ,
\end{eqnarray}
\end{subequations}
 and the integration has been completed. This model is expanding, accelerating and shearing. It is interesting to observe that if we set 
 \[ d=\frac{c^2 b^2 \exp\left( {\frac{(1-3 c b) (e+k)}{1+c b}}\right)  }{(1+c b) (3 c b-1)k^{\frac{2 c b}{1+c b}}}\]
  then the shear vanishes.
 
 As a second example we set 
 \be
 h(x)=(e+kx)^n,
 \ee
  and the integration in \eqref{fff1} can be performed.
 Then the potential functions have the form
 \begin{subequations}\label{fff2}
\begin{eqnarray}
A&=&\frac{c(k n)^{\frac{1}{1+c b}} (e+k x)^{\frac{3 n-cbn-2}{2+2 c b}}}{\beta (t)  \left(d+\frac{c^2 b^2 n\text{  }\left((e+k x)^{\frac{c b (2-3 n)+n}{1+c b}}-(e+k)^{\frac{c b (2-3 n)+n}{1+c b}}\right)}{(1+c b) (-n+c b (-2+3 n))(k n)^{\frac{2 c b}{1+c b}}}\right)^{1/2} },\\
B&=&\frac{\beta (t)}{c \alpha (r)}A, \quad Y=(e+k x)^n.
\end{eqnarray}
\end{subequations}
The solution \eqref{fff2} has interesting features which we will consider later. For now, we note that this model has nonvanishing expansion, acceleration and shear.

It is interesting to note that our solution contains those of Euclidean stars. 
In Euclidean stars the areal  and proper radii are equal; this approach of modeling stars in general  relativity with shear was developed by Herrera and Santos \cite{G5}.  Govender et al.  \cite{G6}  found other particular  Euclidean solutions. 
 
 The metric \eqref{fff1}  will yield the Euclidean star formulation  provided  $B=Y_r$. This yields the condition 
 \be \label{ref}
h'=\frac{h^{\frac{1-c b}{2 (cb+1)}} h'^{\frac{1}{cb+1}}}{\left( d - \frac{c^2 b^2 }{ (cb+1)^2 }\int_1^x  h(z)^{\frac{-2c b}{cb+1}}h'(z)^{\frac{1-cb}{cb+1}} \, dz\right)^{1/2}},
\ee
  If we now take \eqref{fff2} and set 
\begin{eqnarray}
&& b=1,\quad e=0,\quad k^n=\tilde{k}, \nonumber \\
&& d=\frac{c^2  \tilde{k}^{\frac{1-3 c }{1+c }} n^{\frac{1-c }{1+c }}}{(1+c ) (3 c  n-2 c -n)},\quad c=\frac{1-n\pm\sqrt{1-4 n+3 n^2}}{2 (n-1)},
 \end{eqnarray}
then we obtain
\begin{subequations}\label{euc}
\begin{eqnarray}
A&=&\frac{x^{n-1}}{\beta(t)},\\
B&=&\tilde{k}n\frac{x^{n-1}}{\alpha(r)},\\
Y&=&\tilde{k}x^n,\quad \text{where } x= \int\frac{dt}{\beta (t)}+\int\frac{dr}{\alpha (r)}.
\end{eqnarray}
\end{subequations}
This particular solution satisfies  the master equation \eqref{c51} provided that the constants  $n$ and $\tilde{k}$  are related via 
\be \label{ggg1}
 \tilde{k}^2 n^3   - 2 \tilde{k} n^2+ 2n (\tilde{k}-1 ) + 2=0.
 \ee
The solution \eqref{euc} was previously obtained  by Govinder and Govender \cite{G7} in their study of the   junction condition  of an Euclidean star.

As the particular Euclidean star model \eqref{euc} is contained in the more general class of solutions \eqref{fff1} we label the gravitational potentials in \eqref{fff1} as  generalized Euclidean stars.  We note that there may be other special parameter values  for which $h$ can be found explicitly which will correspond  to  other Euclidean stars.

 \section{Equation of state}\label{physi}
 We now study the physical features of the new   generalized  Euclidean model \eqref{fff1} that contains previously obtained  solutions for Euclidean stars as a special case.
 
 The kinematical quantities, the acceleration, the expansion scale   and the magnitude of the shear scalar become
  \begin{subequations}
 \begin{eqnarray}
\dot{u}^a&=&\left(0,\frac{\alpha (r)}{2 (1+c b)^3 h'} \left[\frac{c^2 b^2 (1+c b)}{h} \right.\right. \nonumber \\
&& \left.\left. -\frac{  \varphi (x) }{h^{\frac{2}{1+c b}}  h'^{\frac{2}{1+c b}}} \left((c b-1) h'^2-2 hh''\right)\right],0,0 \right), \\
\Theta &=& \frac{c^2 b^2 (1+c b) hh'^{\frac{2+c b}{1+c b}}+ \varphi (x)h^{\frac{2 c b}{1+c b}}  h'^{\frac{c b}{1+c b}} \left((5+3 c b) h'^2+2 hh''\right)}{2 c b (1+c b)^2 \sqrt{ \varphi (x)}h^{\frac{7+5c b}{2+2c b}} h'^2}, \\
\sigma &=&\frac{ \varphi (x)h^{\frac{2 c b}{1+c b}}  h'^{\frac{c b}{1+c b}} \left((1+3 c b) h'^2-2 hh''\right))-c^2 b^2 (1+c b) hh'^{\frac{2+c b}{1+c b}}}{6c b (1+c b)^2\sqrt{ \varphi (x)}h^{\frac{7+5c b}{2+2c b}}   h'^2},
\end{eqnarray} 
\end{subequations}
respectively, where  we have set 
\[ \varphi (x) \equiv  (1+c b)^2 d-c^2 b^2 \int_1^{x} h(z)^{-\frac{2 c b}{1+c b}} h'(z)^{\frac{1-c b}{1+c b}} \, dz.\]  It is clear that these quantities are nonzero in general.

 The matter variables \eqref{145} become
 \begin{subequations}\label{matter}
\begin{eqnarray}
\mu&=& \left[ 2 \left(\left(1-b^3 c^3\right)  \varphi (x) h^{\frac{bc-1}{1+b c}}\text{  }\left(h'^2+h h''\right) \right. \right. \nonumber \\
&& \left. \left. +b^2 c^2(1+b c) (1+b c (1+b c))h'^{\frac{2}{1+b c}}\right) \right] \left[ b^2 c^2(1+b c)^3 h^2h'^{\frac{2}{1+b c}} \right]^{-1},\\
p_\parallel&=&q,\nonumber\\
&=& \frac{2 \left((b c-1)  \varphi (x) h^{\frac{bc-1}{1+b c}}\text{  }\left(h'^2+h h''\right)-b^2 c^2 (1+b c)h'^{\frac{2}{1+b c}}\right)}{b c (1+b c)^3 h^2h'^{\frac{2}{1+b c}}},\\
p_\perp&=& \frac{(b c-1)}{2 b^2 c^2 (1+b c)^2  \varphi (x)h^{\frac{3bc + 5}{1+b c}} h'^{\frac{2bc+4}{1+b c}}}\left[b^4 c^4 (1+b c) h^{\frac{4}{1+b c}} h'^{\frac{4}{1+b c}}\right.\nonumber\\
&&\left. -b^2 c^2  \varphi (x) h^{\frac{bc +3}{1+b c}} h'^{\frac{2}{1+b c}} \left(2 b c h'^2+(b c-1) h h''\right)+\varphi ^2(x) h^2 \left((b c-1) h'^4\right.\right.\nonumber\\
&&\left.\left. +(3+b c) h h'^2 h''-2 h^2 h''^2+2 h^2 h' h^{(3)}\right)\right],
\end{eqnarray}
\end{subequations}
for  the potentials \eqref{fff1}. Note that the radial pressure and the heat flux are related by \eqref{santo} so that $p_\parallel=q$ in our solution. For nonzero tangential pressure $p_\perp$ we must have $bc\ne 1$. If we set $bc=1$  then $p_\perp =0$ and the heat flux becomes negative.

We observe from \eqref{matter} that the relationship
 \be \label{muu} 
 p_\parallel (\mu)=\lambda \mu, \quad \lambda =-\frac{cb}{c^2b^2 +cb +1},
 \ee
  is satisfied. Hence the generalized Euclidean star model \eqref{fff1} always satisfies a linear barotropic equation of state. This is a result independent of the analytic form of the arbitrary function $h(x)$. The Lie theory of differential equations has produced a family of exact solutions for general relativistic stellar models which is characterized by an equation of state.    If we use the forms   \eqref{euc} of the Govinder and Govender \cite{G7} particular solution then it is easy to show that 
 \be \label{euces}
  p_\parallel (\mu)=\lambda \mu, \quad \lambda = \frac{\tilde{k}^2 n^3-2 n+2}{\tilde{k}^2 n^2 (2-3 n)},
  \ee
  so that \eqref{muu} is satisfied. Consequently their  solution  is a particular case of our more general case.   The thermodynamic properties  and other physical features of \eqref{euces} point to a physical reasonable model. In particular the causal   temperature is higher than that of the  noncausal temperature in the core of the star.

\section{Energy conditions\label{energy}}  
  It is necessary to choose a particular form of the metric to study further physical features. Consequently  we set  $b=1$, $e=0$, $k^n=\tilde{k}$,  $d=\frac{c^2  \tilde{k}^{\frac{1-3 c }{1+c }} n^{\frac{1-c }{1+c }}}{(1+c ) (3 c  n-2 c -n)}$ and  $c=-3$   in  \eqref{fff2}. This   yields  the  kinematical quantities
\begin{subequations}
\bq
 \dot{u}^a&=&\left(0,\frac{9 (n-1)\text{  }\alpha (r)}{4 \tilde{k}^2 n (5 n-3)x^{2 n-1}},0,0 \right),\\
\Theta &=&\frac{(1-3 n) }{2 \tilde{k} \sqrt{n (5 n-3)}x^{n}},\\
\sigma &=&-\frac{1}{6\tilde{k} \sqrt{n (5 n-3)}x^n}.
\eq
\end{subequations}
The kinematical  quantities are well defined if $n>3/5$.

The dynamical quantities have the form
\begin{subequations}\label{39}
\bq
\mu&=&\frac{7 (n-1) }{2\tilde{k}^2 (5n-3)x^{2 n}}\label{amzi1},\\
p_{\parallel}&=&q,\nonumber\\
&=&\frac{3 (n-1) }{\tilde{k}^2 (5n-3)x^{2 n}},\\
p_{\perp}&=&\frac{2 (n-1)^2 }{\tilde{k}^2 n (5n-3)x^{2 n}}\label{amzi2},
\eq
\end{subequations}
in our example.
From \eqref{39} we note that $\mu >0$, $p_{\parallel}>0$, $p_{\perp}>0$ for all  $n>1$. 

The energy conditions for a matter distribution with isotropic pressures in the presence of heat flux were defined by Kolassis et al. \cite{1k}.  This was extended by Chan \cite{H4} for anisotropic pressures with heat flow. We follow the approach of Chan \cite{H4} when evaluating the energy conditions. Again we note that $p_{\parallel}=q$ is a restriction that applies in our model; for a general heat conducting fluid this is not true.  For our example we evaluate the quantities
\bq
&& |\mu+ p_{\parallel}| -2|q |=\frac{2 (n-1) }{\tilde{k}^2 (5n-3)x^{2 n}},\\
&& \mu - p_{\parallel} + 2p_{\perp} + \sqrt{(\mu+ p_{\parallel} )^2 - 4 q^2}=\frac{2 (n-1) (5 n-2) }{\tilde{k}^2 n (5n-3)x^{2 n}}.
\eq
 We observe that these quantities are nonnegative if $n\ge1$. In addition we have the following:\\
\noindent (i) weak energy conditions:
\bq \label{weak}
 E_{\text{wec}}&=&\mu- p_{\parallel} + \sqrt{(\mu+ p_{\parallel} )^2 - 4 q^2}\nonumber\\
 &=&\frac{6 (n-1) }{\tilde{k}^2 (5n-3)x^{2 n}}, 
 \eq
(ii)  dominant energy conditions:
\bq 
E^{(1)}_{\text{dec}}&=&\mu - p_{\parallel}\nonumber\\
&=&\frac{2 (n-1) }{\tilde{k}^2 (5n-3)x^{2 n}}, \\
E^{(2)}_{\text{dec}}&= &\mu - p_{\parallel} - 2p_{\perp} + \sqrt{(\mu+ p_{\parallel} )^2 - 4 q^2}\nonumber\\
&=& \frac{2 (n-1) (2+n)}{\tilde{k}^2 n (5n-3)x^{2 n}},
 \eq
(iii) strong energy conditions:
\bq
 E_{\text{sec}}&=&2p_{\perp} + \sqrt{(\mu+ p_{\parallel} )^2 - 4 q^2}\nonumber\\
 & =&\frac{4 (n-1) (2 n-1) }{\tilde{k}^2 n (5 n-3)x^{2n}} .
 \eq
 We observe   that $E_{\text{wec}}\geq 0$, $E^{(1)}_{\text{dec}}\geq 0$, $E^{(2)}_{\text{dec}}\geq 0$, and $ E_{\text{sec}}\geq 0$ when $n\geq 1$. Hence the weak, dominant and strong energy conditions are satisfied in this example. This indicates that the matter distribution
in generalized Euclidean star is physically reasonable. Note that Govinder and Govender \cite{G7} also showed  that their Euclidean star model \eqref{euc},  which is a special case of our solution \eqref{fff1}, satisfies  these conditions when 
 $
 \tilde{k}=-2/9, n=3 $ and $x= -(c_1r^2+c_2t^2)
 $ 
 where $c_1$ and $c_2$ are positive constants.

 \section{Discussion}\label{5sec4}
We have shown  that the Lie symmetry approach to differential equations is a useful tool that can assist in the  search of exact solutions to the junction condition  relating the radial pressure with the heat flux at the boundary of relativistic radiating star. Using the Lie approach we generate several new exact solutions to the boundary condition equation. Our solutions  regain  previously known  solutions for the Euclidean star as a special case. The model of Govinder and Govender \cite{G7} arises as a special case; our analysis shows that their results are part of a more general class invariant under  the action of Lie symmetry generators. A noteworthy feature of our new generalized Euclidean star models is that they  obey a linear barotropic equation of state in general. Most of the radiating stellar models found in the past do not share this feature. The analysis of the weak, dominant and strong  energy conditions in the example  shows that the matter distribution is physically acceptable.

\begin{acknowledgements}
GZA and KSG thank the  University of KwaZulu-Natal and National Research Foundation  for continuing support.
SDM acknowledges that this work is based upon research supported by the South African Research Chair Initiative of the Department of Science and Technology and the National Research Foundation.\end{acknowledgements}

\section{Appendix\label{Appen}}
In the Appendix we consider the Lie symmetries \eqref{mar7} and the various cases for which the boundary condition \eqref{c51} can be integrated. The case of generalized Euclidean stars was considered in Sect. \ref{xzx}. The other integrable cases are given below.

\subsection{Static solution: Generator $X_1$\label{secc2}}
Using the generator
\be 
X_1=A\frac{d\beta }{dt}\frac{\partial}{\partial A}-\beta (t)\frac{\partial}{\partial t},
\ee
 we determine the surface conditions
\begin{equation}
\frac{dt}{-\beta (t)}=\frac{dr}{0}=\frac{dA}{A\frac{d\beta }{dt}}=\frac{dB}{0}=\frac{dY}{0}.
\end{equation}
The invariants are given by $r$ and
\be\label{inv51}
A=\frac{f(r)}{\beta (t)},\quad
B=g(r),\quad
Y=h(r).
\ee
 Note that for the solution corresponding to \eqref{inv51} the gravitational potentials become static and the star is not radiating. 

\subsection{Shear-free solution: Generator $aX_1+X_3$}
 Using the generator
\be 
aX_1+X_3=\left( 1+a\frac{d\beta }{dt}\right) A\frac{\partial}{\partial A}+B\frac{\partial}{\partial B}+Y\frac{\partial}{\partial Y}-a\beta (t)\frac{\partial}{\partial t},
\ee
we determine   the  surface conditions
\begin{equation}
\frac{dt}{-a\beta (t)}=\frac{dr}{0}=\frac{dA}{\left( 1+a\frac{d\beta }{dt}\right) A}=\frac{dB}{B}=\frac{dY}{Y}.
\end{equation}
The invariants are given by  $r$ and 
\begin{subequations}\label{inv53}
\begin{eqnarray}
A=\frac{ f(r)}{\beta (t)\exp\left( \int\frac{dt}{a\beta (t)}\right) }, \quad
B=\frac{g(r)}{\exp\left( \int\frac{dt}{a\beta (t)}\right) }, \quad
Y=\frac{h(r)}{\exp\left( \int\frac{dt}{a\beta (t)}\right) }.
\end{eqnarray}
\end{subequations}
In this case the shear vanishes. 

\subsection{Generator   $X_2-bX_1$: Linear equation}
If we set 
\be 
\frac{b^2 f^3-2 h f' h'+f \left(h'^2+2 h h''\right)}{2 b f^2 h h'}=0,
\ee
 in \eqref{ss2}  then we integrate to find
\be\label{nnnew23}
f(x)=\frac{\sqrt{h} h'}{\sqrt{m-b^2 h}},
 \ee
 where $m$ is a constant.  Substituting \eqref{nnnew23} into   \eqref{ss2} we have
 \be \label{linear1}
 g'+\left(\frac{m h'}{2 m h-2 b^2 h^2}\right)g -\left(\frac{b \left(\left(b^2 h-2 m\right) h'^2+2 h \left(b^2 h-m\right) h''\right)}{2 \sqrt{h} \left(m-b^2 h\right)^{3/2}}\right)=0,
 \ee
 which is now linear in $g$.
 Equation \eqref{linear1} can be integrated to give
 \begin{eqnarray}  g(x)&=&\sqrt{\frac{m-b^2 h} {h}} \times \nonumber \\
 &&  \left(\int_1^{x } \frac{b \left[\left(b^2 h(z)-2 m\right) h'(z)^2+2 h(z) \left(b^2 h(z)-m\right) h''(z)\right]}{2 \left(m-b^2 h(z)\right)^2} \, dz+n\right),  \label{nnnew1}
\end{eqnarray}
where $n$ is arbitrary constant.

The potential functions can be written as 
\begin{subequations}\label{matric1}
\begin{eqnarray}
A&=&\frac{ h'(x)}{\beta (t)}\sqrt{\frac{h(x) }{m-b^2 h(x)}},\\
B&=&\frac{ 1}{\alpha (r)}\sqrt{\frac{m-b^2 h(x) }{ h(x)}} \times \nonumber\\
&& \left(\int_1^{x } \frac{b \left[\left(b^2 h(z)-2 m\right) h'(z)^2+2 h(z) \left(b^2 h(z)-m\right) h''(z)\right]}{2 \left(m-b^2 h(z)\right)^2}  dz+n\right),\\
Y&=&h(x),
\end{eqnarray}
\end{subequations}
 which is another particular solution  to the master equation \eqref{c51}. Note that $h$ is an arbitrary function of $x= \int\frac{dt}{b\beta (t)}+\int\frac{dr}{\alpha (r)} $  in  this solution.

\subsection{Generator  $X_2-bX_1$: Bernoulli equation}
If we set
 \be \label{ricca1}
 h(x)=\frac{c}{f^2},
 \ee
 where $c$ is arbitrary, \eqref{ss2} reduces to the Bernoulli equation
 \be \label{ricca3}
 g'+\left(\frac{4 f'}{f}-\frac{f''}{f'}\right)g +\left(\frac{b^2 f^8+20 c^2 f'^2-4 c^2 f f''}{4 b c^2 f^2 f'}\right)g^2=0.
 \ee
 Equation \eqref{ricca3} can be integrated in general to give
\be\label{ricca5} 
g(x)=\frac{f'}{f^4 \left(\int_1^x \frac{b^2 f(z)^8+20 c^2 f'(z)^2-4 c^2 f(z) f''(z)}{4 b c^2 f(z)^6} \, dz+w\right)},
\ee
where $w$ is an arbitrary constant.

Then the potential functions have the form
\begin{subequations}
\begin{eqnarray}
A&=&\frac{f(x)}{\beta (t)},\\
B&=&\frac{f'(x)}{\alpha (r)f^4(x) \left(\int_1^x \frac{b^2 f(z)^8+20 c^2 f'(z)^2-4 c^2 f(z) f''(z)}{4 b c^2 f(z)^6} \, dz+w\right)},\\
Y&=&\frac{c}{f^2(x)},
\end{eqnarray}
\end{subequations}
 which is  a new solution to the master equation \eqref{c51}. Observe  that  $f$ is arbitrary function of $x= \int\frac{dt}{b\beta (t)}+\int\frac{dr}{\alpha (r) }$ for this solution.

 \subsection{Generator  $X_2-bX_1$: Riccati equation} 
 While (\ref{ss2}) is a Riccati equation, we observe that setting 
\be
f(x)=c h',
\ee
 where $c$ is constant, simplifies it considerably to
\be\label{riccc}
g'-\left(\frac{1+b^2 c^2}{2 b c h}\right)g^2+\left(\frac{b c h'^2}{2 h}+b c h''\right)=0.
\ee
We still cannot solve \eqref{riccc} in general. However it can be integrated for some particular functional forms of $h$. If we set $h$ to be a simple quadratic function
\be 
h(x)=x^2,
\ee
then \eqref{riccc}  gives
\be
g(x)=\left(\frac{b c x }{1+b^2 c^2}\right)\left[1-\sqrt{9+8 b^2 c^2} \left(\frac{k x^{\sqrt{9+8 b^2 c^2}}-1}{k x^{\sqrt{9+8 b^2 c^2}}+1}\right)\right],
\ee
where $k$ is a constant. Then the potential functions have the form
\begin{subequations}\label{xxx}
\bq
A&=&\frac{2 c}{\beta(t)}  \left(\int \frac{\, dt}{b\beta (t)} +\int \frac{\, dr}{\alpha(r)} \right),\\
B&=&\left(\frac{b c  \left(\int \frac{\, dt}{b\beta (t)} +\int \frac{\, dr}{\alpha(r)} \right) }{\left( 1+b^2 c^2\right) \alpha (r)}\right)\nonumber\\
&&\times\left[1-\sqrt{9+8 b^2 c^2} \left(\frac{k  \left(\int \frac{\, dt}{b\beta (t)} +\int \frac{\, dr}{\alpha(r)} \right)^{\sqrt{9+8 b^2 c^2}}-1}{k \left(\int \frac{\, dt}{b\beta (t)} +\int \frac{\, dr}{\alpha(r)} \right)^{\sqrt{9+8 b^2 c^2}}+1}\right)\right],\\
Y&=& \left(\int \frac{\, dt}{b\beta (t)} +\int \frac{\, dr}{\alpha(r)} \right)^2.
\eq
\end{subequations}
This is another new solution for the master equation \eqref{c51}.

 The solution \eqref {xxx} does not obey an equation of state in general. However when we set the constant $k=0$, then \eqref{xxx} becomes
\begin{subequations}\label{yyy}
\bq
A&=&\frac{2 c}{\beta(t)}  \left(\int \frac{\, dt}{b\beta (t)} +\int \frac{\, dr}{\alpha(r)} \right),\\
B&=&\frac{b c \left(1+\sqrt{9+8 b^2 c^2}\right)}{\left(1+b^2 c^2\right) \alpha (r)} \left(\int \frac{\, dt}{b\beta (t)} +\int \frac{\, dr}{\alpha(r)} \right),\\
Y&=& \left(\int \frac{\, dt}{b\beta (t)} +\int \frac{\, dr}{\alpha(r)} \right)^2.
\eq
\end{subequations}
The special case \eqref{yyy} obeys the  barotropic equation of state
\be 
p_{\parallel}(\mu)=\lambda \mu ,
\ee
where $\lambda= \frac{b^2 c^2 \left(3-\sqrt{9+8 b^2 c^2}\right)-3 \sqrt{9+8 b^2 c^2}+5}{2 \left(7+5 b^2 c^2+b^4 c^4\right)}$.

\end{document}